\documentclass{vgtc}                          

\ifpdf
  \pdfoutput=1\relax                   
  \usepackage{graphicx}                
  \DeclareGraphicsExtensions{.pdf,.png,.jpg,.jpeg} 
\else
  \usepackage{graphicx}                
  \DeclareGraphicsExtensions{.eps}     
\fi%

\graphicspath{{figures/}{pictures/}{images/}{./}} 

\usepackage{microtype}                 
\PassOptionsToPackage{warn}{textcomp}  
\usepackage{textcomp}                  
\usepackage{mathptmx}                  
\usepackage{times}                     
\usepackage{cite}                      
\usepackage{tabu}                      
\usepackage{booktabs}                  
\usepackage{indentfirst}

\onlineid{0}

\vgtccategory{Research}

\vgtcinsertpkg

\title{TangibleChannel: An Innovative Data Physicalization System for Visual Channel Education \vspace{-0.3cm}}

\vspace{-0.5cm}

\author{Siqi Xie\thanks{e-mail: Siqi.Xie18@student.xjtlu.edu.cn}
\and Yu Liu\thanks{e-mail:Yu.Liu02@xjtlu.edu.cn} %
\and Lingyun Yu\thanks{e-mail:Lingyun.Yu@xjtlu.edu.cn}}
\affiliation{\vspace{-0.3cm}\scriptsize School of Advanced Technology, Xi'an Jiaotong-Liverpool University}

\teaser{
\vspace{-0.5cm}
  \centering
 \setlength{\abovecaptionskip}{-0.02cm} 
  \includegraphics[width= 0.7\textwidth]{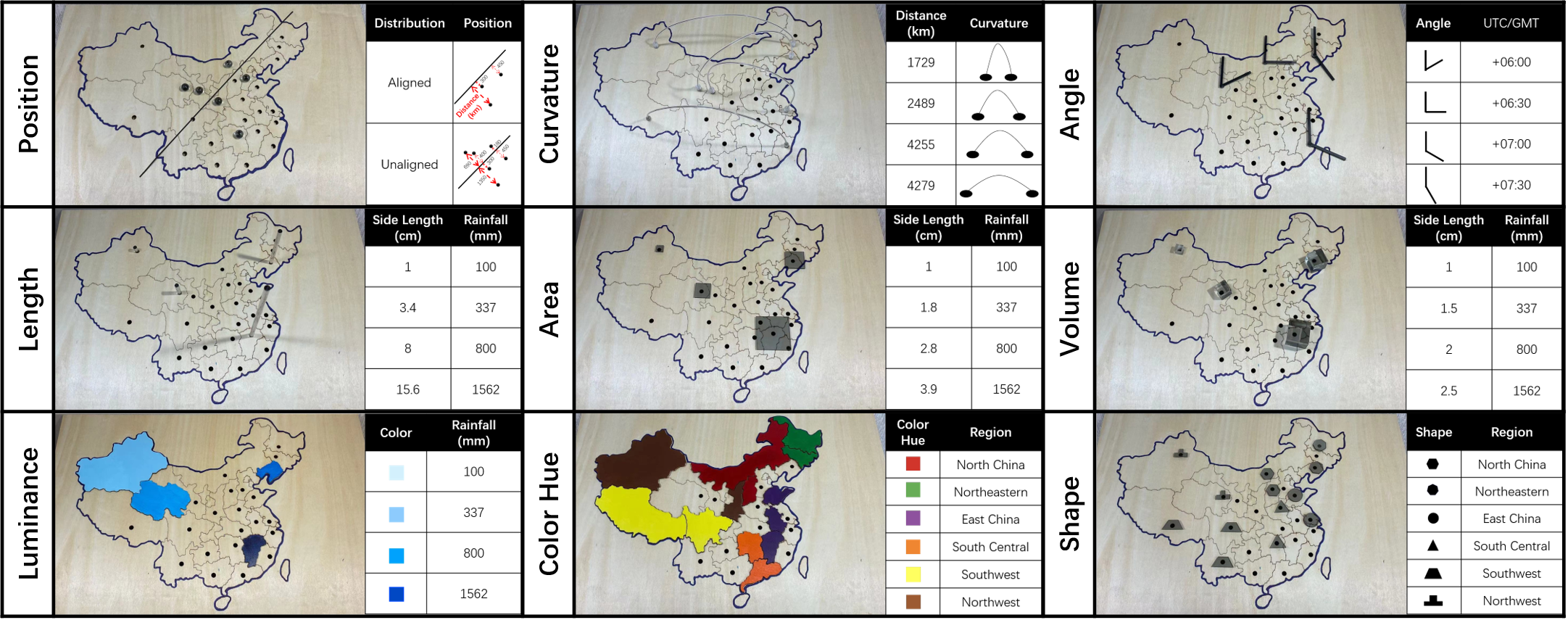}
  \caption{\textit{TangibleChannel}: The data physicalization used for teaching the effective use of visual channels.}
  \label{fig:physical}
  \vspace{-0.2cm}
}

\abstract{In this paper, we provide an overview of our attempts to harness data physicalizations as pedagogical tools for enhancing the understanding of visual channels.  
We first elaborate the research goals that we have crafted for the physicalization prototype, shedding light on the key principles that guided our design choices. 
Then we detail the materials and datasets we employed for nine channels on our physicalization prototype. 
A preliminary pilot study is followed to validate its effectiveness.
In the end, we present our upcoming research initiatives, including a comparative study for assessing the usability of the physicalization system.
In general, the main purpose of our work is to stimulate a wider engagement among visualization educators and researchers, encouraging them to delve into the potentialities of data physicalization as an innovative addition to contemporary teaching methodologies.}

\CCScatlist{
    \CCScatTwelve{Human-centered computing}{Visualization}{}{}
    \vspace{-0.13cm}
}

\nocopyrightspace

\begin{document}
\maketitle

\section{Introduction}

\vspace{-0.13cm}

Data physicalization involves the conversion of digital data into tangible forms, allowing users to perceive and interact with the data through multiple human senses, mainly including tactility and audition \cite{dragicevic2020dat}, achieved by using physical objects and materials \cite{jansen2013evaluating}. 
The main purpose of data physicalization is to facilitate the straightforward and direct exploration, interpretation, and communication of data \cite{jansen2013evaluating}. 
Because of its directness, physicalization has proven effective in educational contexts, particularly in enhancing students' comprehension of data structures and algorithms. This approach fosters the creation of accessible visualizations and empowers students to communicate outcomes more accurately \cite{van2022data}. 
Moreover, data physicalization has the potential to engage students in immersive learning experiences that encourage critical thinking. The applications of data physicalization are extensive and diverse, spanning fields from geospatial education and architectural modeling to STEM learning \cite{djavaherpour2017physical, hayes2018exploring}. 



Despite its various applications, the utilization of data physicalization in supporting visualization education remains relatively under-explored. This observation has spearheaded the development of a new research agenda, aimed at bolstering the teaching of visual channels in visualization through data physicalization. 
Understanding visual coding is fundamental when starting to delve into visualization, however, many beginners fail to grasp its comprehensive and profound nature. 
To address this issue, we propose \textit{TangibleChannel} to foster a more intuitive and interactive learning experience than traditional teaching methods. Our system is designed based on five key design factors and three overarching research goals, all of which will be introduced in the sections that follow.




\vspace{-0.2cm}
\section{Related Work}

\vspace{-0.13cm}


Research in data physicalization has revealed its usefulness in visualizing various data, such as personal information, and its impact on domestic behaviors, fostering personal reflection \cite{steger2022ecorbis}. Moreover, its potential as an educational tool has been acknowledged, being used in subjects as diverse as organic chemistry, thermodynamics, seismology, insect ecology, and astronomy\cite{physlist}. Interactive tools like Data Bricks Space Mission \cite{ambrosini2022data} makes astronomical data tangible using Lego bricks. 
Building upon these existing studies, our aim is to investigate the transformative potential of data physicalization within learning environments, with a specific focus on its capacity to elevate students' proficiency in data visualization.

\vspace{-0.2cm}
\section{Design}

\vspace{-0.13cm}


For designing the physical education system, we have first developed a basic understanding of creating physical representations. 
This involves a design process that includes thoughtful selections of suitable materials for visual channels, as well as a deep comprehension of their unique strengths and limitations (Research Goal 1).
Simultaneously, our research seeks to uncover the potential advantages of integrating physical visualization into educational contexts (Research Goal 2). In particular, we aim to explore whether the value of data physicalization primarily stems from its tangible form, which can be observed, or from the direct interaction between users and physical objects (Research Goal 3).
To conceptualize and establish our data physicalization prototype, we incorporate the following five design considerations:\\
\textbf{1. Pedagogical Objectives}: By collaborating with experienced visualization educators and conducting comprehensive reviews of relevant online courses, we have gained valuable insights into our teaching objectives. Our primary focus is to enhance the learning process by facilitating an in-depth comprehension of selecting appropriate visual channels for distinct data types. \\
\noindent\textbf{2. Target Audience}: Our design process for physicalization is carefully tailored to our target audience, which mainly includes students engaged in the study of visual channels and educators in the visualization domain. \\
\noindent\textbf{3. Data Quality}: We ensure that the data incorporated in our physicalization process is both accurate and reliable. Our data is all obtained from reliable public data sites.\\
\noindent\textbf{4. Interactivity}: Visualization processes are inherently interactive. Thus, our physicalization system is designed to foster interactivity, allowing for exploration, comparison, and hands-on manipulation of these physical channels. This approach is aimed at promoting student engagement and facilitating in-depth understanding.\\
\noindent\textbf{5. Aesthetic Appeal}: We strive to create visually pleasing and harmonious representations that integrate elements such as color, shape, and texture. Our intention is not only to enhance the learning experience but also to captivate and inspire learners through the sheer beauty of the design.

Figure \ref{fig:physical} showcases our TangibleChannel system prototype, crafted by laser-engraving the basic outlines of Chinese provinces onto a wooden board. To instantiate the coded channels, we employed a diverse range of epoxy resin materials, including balls, blocks, sticks, tubes, tokens, and wooden shades, representing the data of a distinct province.
To enhance the interactivity of this educational tool, we embedded magnets in the wood panels. These magnets attract corresponding wooden blocks, which are outfitted with magnetic elements. Additionally, wooden sticks are securely glued beneath each epoxy component, designed to be inserted into various pre-drilled holes in the wooden board, allowing for flexible and engaging manipulation.
We opted for a light grey hue for the epoxy resin. This choice ensures that the map's details remain clearly visible and easy to read, striking a balance between complete opacity (as with wood) and full transparency. 
In selecting the datasets to be represented by different channels, we include both categorical and quantitative data. 
Meanwhile, we intentionally chose datasets that allow the various encoding channels to showcase both large and small differences between data points. This design decision was made with the aim of creating a conducive learning environment in which students can clearly discern and understand the differences between each channel, thereby gaining a deeper appreciation of how information is presented.

\vspace{-0.05cm}

\textbf{Position} is represented by the distance of the ball from the Heihe-Tengchong line\footnote{Wikipedia. Heihe–Tengchong Line, 2022}, devided into two categories: same-side and opposite-side. This distinction aids users in discerning the accuracy of position perception in both aligned (points on the same side) and unaligned (points on different sides) contexts.
\textbf{Curvature} represents the distance between different cities, achieved by inserting pipes into holes at both ends. Greater distances resulted in a more pronounced curvature of the pipe.
\textbf{Angle} between two gray sticks suggests the timezone difference between the province and UTC, inspired by the clock's mechanism where angles represent time.
The following four visual channels depict the annual rainfall data for different provinces in China\footnote{http://www.stats.gov.cn/sj/ndsj/2011/indexeh.htm}.
\textbf{Length} is portrayed by gray epoxy sticks of varying lengths. Each unit length of the stick (1 cm) represented 100mm of annual precipitation.
\textbf{Area} is indicated by square tokens beginning from $1 cm^2$. The square of the length of sides corresponds to the precipitation.
\textbf{Volume} is represented using cubes with varying side lengths beginning from $1 cm^3$, and the difference in volume encodes the different amount of precipitation. .
\textbf{Color Luminance} incorporates a blue hue.
Each luminance corresponds to a different rainfall level—darker hues signify heavier rainfall and vice versa. 
The visual channels of \textbf{Color Hue} and \textbf{Shape} relays the general region classification of Chinese provinces\footnote{Wikipedia. Administrative divisions of China, 2023}.
For \textbf{Shape}, we deliberately selected six distinct shapes. Among these, three shapes—hexagon, heptagon, and circle—are closely related and may introduce a degree of ambiguity, thereby serving to emphasize the potential confusion that may arise when shapes are too similar. On the other hand, the remaining three shapes—triangle, trapezoid, and convex shape—are markedly different from each other, which aids users in more readily identifying and comprehending the distinct data categories each shape represents. This selection was strategically designed to illustrate the significance of shape choices in clearly and effectively conveying disparate categories of information.
\textbf{Color Hue} is chosen from the ColorBrewer palette\footnote{https://colorbrewer2.org/} following a similar strategy with shape. We have chosen a set of similar hues to illustrate the potential for ambiguity when colors are too closely related, as well as a set of distinctly contrasting hues to emphasize the ease of interpretation when colors are clearly distinguishable, thereby highlighting the importance of thoughtful hue selection in effective data communication..



The \textit{TangibleChannel} prototype offers a unique educational platform for students and teachers to explore physical channels. To ascertain the initial impact of the prototype on improving students' understanding of channel attributes, we conducted a pilot study in an Information Visualization course involving 24 master students. The prototype was used as a supplementary tool for teaching materials. This pilot study showed its effectiveness in helping students comprehend unique channel characteristics and distinguish between various channels, especially within 3D context. Here, the advantages and drawbacks of employing physical objects to represent diverse visual channels became more evident.

\vspace{-0.2cm}
\section{Discussion and Conclusion}

\vspace{-0.1cm}

In this study, we have introduced a physicalization prototype of an educational system aimed at imparting knowledge about visual channels. This system, built with tangible materials such as wood and epoxy resin, consists of nine physical channels designed and fabricated to bolster student engagement and comprehension of visual coding concepts. 
Our aspiration is that the insights and discoveries will inspire and inform educators and visualization scholars to delve into the potential of data physicalization in their teaching practices. Therefore, as part of our future research, we plan to explore the learning outcomes facilitated by our physicalization prototype compared to a screen-based education system. This future research will further contribute to the understanding of data physicalization as an additional tool for visual coding education.


\vspace{-0.15cm}
\section*{acknowledgments}
\vspace{-0.12cm}
The work was partially supported by NSFC (62272396) and XJTLU RDF (22-01-092, 19-02-11).
\vspace{-0.15cm}


\bibliographystyle{abbrv-doi}

\bibliography{ref/ref}

\end{document}